# Bonaparte and the astronomers of Brera Observatory

Elio Antonello
*INAF-Osservatorio Astronomico di Brera*
elio.antonello@brera.inaf.it

**Abstract.** In Northern Italy, between 1796 and 1814, Napoleon Bonaparte formed a Republic, and then a Kingdom, controlled by France. Milan was the capital of the State, and the Brera Palace was its main cultural centre, as regards both the arts and the sciences. Bonaparte probably intended to strengthen this characteristic of Brera, aiming at increasing its Italian and European relevance. We will discuss in detail in which way he interacted with the astronomers of the Brera Observatory, and in particular with Barnaba Oriani, that was considered the local main representative of the *république des lettres*, i.e. the world of literature, arts and sciences. We propose a possible reconstruction of the effects of those complicated historical events on the Italian astronomy and on its relations with the European one.

## 1. Introduction

After Galileo's discoveries, several Italian scholars gave significant contributions to the observational astronomical researches during the $17^{th}$ and $18^{th}$ century, in particular in Bologna and Rome, that is in the Papal States. The official culture, however, was against the Copernican theory, or at most it assumed the heliocentrism just as an hypothesis, waiting for an experimental confirmation of the motion of the Earth. The Bologna Observatory had been the main astronomical institution in Italy until the second half of the $18^{th}$ century, when the Brera Observatory in Milan had become progressively the reference point for Italian astronomy. In the present note we will summarize the first decades of the Brera Observatory, and then we will try to illustrate in some detail the events occurred after 1796, when Napoleon Bonaparte formed a State in Northern Italy, controlled by France. It was a Republic, and then a Kingdom, that included initially only part of Lombardy and Emilia-Romagna, and then also Veneto, Trentino, Friuli and Marche. The main universities of the Kingdom were Bologna, Padua, Pavia, Modena, Ferrara and Macerata, but there were other high schools at university level, such as those of Brera (the previous *Scuole Palatine* of Milan). Brera Palace was the main cultural institution as regards both the arts and the sciences in the capital of the State, Milan, and its Observatory was for some time the only public scientific institution in the town. Bonaparte probably intended to strengthen this characteristic of Brera, aiming at increasing its Italian and European relevance. We discuss in detail in which way he interacted with the astronomers of the Observatory, and in particular with Barnaba Oriani, that was considered the local main representative of the 'république des lettres', i.e. the world of literature, arts and sciences. We used both unpublished archive material preserved in the Observatory and poorly known published papers to attempt to clarify some points concerning the relation between Bonaparte and Oriani.

## 2. The Brera Observatory[1]

The Duchy of Milan had been an independent State from 1395 to 1499, and then it was ruled by foreign powers, in particular by the Austrian Habsburg house from about 1706. Maria Theresa (from 1740 to 1780) and her son Joseph II (from 1780 to 1790) carried out a series of substantial reforms that allowed the development of the State, which enjoyed a peaceful period of about fifty years until the Napoleonic wars. There were progresses in many fields: economy, government, education, science, arts and culture. It was the period of the Milanese Enlightenment (see for instance Israel, 2011, pp. 356-364), and probably it was not by chance that even astronomy took advantage of that favourable situation.

Regular astronomical and meteorological observations were organized in the Collegium of Jesuits of Brera, the present day Brera Palace[2], at the end of 1762, after the move of Father Louis Lagrange (1711-1783)[3] from the Marseille Observatory to Milan. In 1763 Father Ruggiero G. Boscovich (1711-1787) was appointed at the chair of mathematics of the University of Pavia, and, during a short stay in Brera in 1764, he was informed of the idea to build a new Observatory[4]. The Jesuit Boscovich was a polymath, and he had some experience also in engineering, so he was entrusted with the project. Between the second half of 1764 and the beginning of 1765 he designed the structure. The construction began in April 1765, and six months later it was essentially completed (Proverbio, 1997). The Observatory, or Specola, became quickly renowned in Europe (Lalande, 1776, p. 596; Lalande, 1790, I p. 303, V p. 435)[5]. Schiaparelli (1938), in a posthumous publication, wrote that if Boscovich had been able to realize all his ideas, and if he had been fairly and strongly supported, Brera could have been one of the main observatories in Europe, or perhaps the most important one, at least on the Continent. But the human passions prevented such a good outcome. Unfortunately, there was an increasing disagreement between Boscovich and other Jesuits, such as Lagrange, and part of the difficulties probably derived from Boscovich's character.

---

[1] The history of the Observatory may be found in some papers and books (in Italian): Schiaparelli (1880), Zagar (1963); Tagliaferri et al. (1983); see also Antonello (2010).
[2] The Brera Palace today hosts the *Pinacoteca di Brera* (Art Gallery, from 1776), the *Accademia* (School of fine arts, from 1776), the National Library *Braidense* (from 1773), the Botanical Garden (from 1774), the Astronomical Observatory (Specola) and the Museum of astronomical instruments. A close by Palace hosts the *Istituto Lombardo Accademia di Scienze e Lettere* (in the Brera Palace from 1810 to 1959).
[3] There is no relation with the famous Giuseppe Luigi (Joseph Louis) Lagrange, from Turin.
[4] For a short biography and a description of the relation between Boscovich and the Brera Observatory, see Antonello (2013). For his detailed biography, see Hill (1961).
[5] See for example Burney (1773; pp. 89-93) for an enthusiastic opinion about Boscovich and the Observatory.

At the end of 1771 the chancellor Kaunitz wrote to his minister plenipotentiary in Milan, Firmian, a letter where objections were raised concerning the Observatory, and the preparation of a development plan was requested. We may presume that Kaunitz intended to satisfy the needs of a serious scientific research in an Observatory that had to be of public utility and not just a private institute[6]. His criticism therefore had to be intended in a constructive sense, but Boscovich was upset. In February 1772 he sent to Firmian the requested plan as a part of a very detailed document, the *Risposta* to a paragraph in a letter by Prince Kaunitz, which is very useful for the history of the Observatory (Proverbio 1987); it gives information on the research activity and its organization, and it includes a description of the Specola and of the rich astronomical instrumentation. A lively exchange of letters between Boscovich and Firmian then ensued. The Jesuit complained bitterly at the way the authorities had criticized his activity, and since he considered unsatisfactory Kaunitz's proposals and Firmian's replies, he renounced the position of astronomer and the professorship. His resignation was accepted at the beginning of 1773. In August, Pope Clement XIV suppressed the Society of Jesus, and Boscovich moved to France, where he was appointed director of the naval optics of the French Marine. He however continued to keep in contact with his colleagues in Milan, and when he came back to Italy, he spent some months near Brera in 1786, before the definitive deterioration in his health.

Many of Boscovich's proposals contained in the *Risposta* were accepted by Kaunitz, and were included in the final development plan for the Observatory. It must be emphasized the thoughtfulness and care of Vienna Court for the development of the Specola, until the Napoleonic wars. For example, the optics for an achromatic telescope were purchased in England (Dollond), and the telescope mounting was realized by Giuseppe Megele, an engineer who was moved from Vienna to Milan. Also an equatorial telescope of Sisson was purchased, and it was used (with some improvements) for almost a century[7].

Those decades saw the development of the personality of the priest Barnaba Oriani (1752-1832). Paolo Frisi (1728-1784), professor at the *Scuole Palatine*[8], had introduced him to theoretical astronomy, and then Oriani mastered the practical astronomy in the Observatory. Such an education enabled him to achieve results that would have made him famous in Europe, such as the calculation of the orbit of Uranus, discovered by William Herschel (1738-1822) in 1781. He published his observational and theoretical results about Uranus in the annual *Ephemerides* of Milan starting from 1785, with the final theory in 1793, and further refinements until 1814 (Bianchi, 1933). In 1786 he got funds from the Government for a journey of a few months to visit some European research institutions, and with the task of interacting in London with Ramsden, to discuss the details of a new large mural quadrant (with radius of 2.44 m)[9]. The instrument was installed in Brera in 1791, and the subsequent year a telescope of 18 cm was ordered to Herschel.

---

[6] The suppression of Jesuits was expected, and therefore the Government was already taking care of the Brera institutions.
[7] With this instrument Schiaparelli discovered the asteroid Esperia in 1861.
[8] Frisi, who gave contributions also to the studies of the figure of the Earth, was in charge in particular of the education of engineers in Milan.

## 3. Geodetic measurements

By the end of 1777, a renowned geographer and cartographer, Antonio Rizzi-Zannoni (1736-1814) from Padua (Republic of Venice), proposed a project to Kaunitz that included the surveying of the Milanese territory[10]. In fact, the existing topographic map of the Duchy of Milan was considered insufficient. Kaunitz forwarded the proposal to Firmian in order to get the opinion of the astronomers of Brera and of Frisi. The idea was that the astronomers and Frisi had to deal with the scientific part of the project, and Rizzi-Zannoni with the cartography. After the retirement of Father Lagrange, the astronomers in Brera were the Father Francesco Reggio (1743-1804), Father Angelo Cesaris (or De Cesaris, 1749-1832, Boscovich's pupil), and Oriani. The astronomers had already made some astronomical-geodetic measurements, stimulated by previous requests by Cassini III de Thury, and they deemed too large a project that of Rizzi-Zannoni, but they approved it at least in principle. Their main remark, however, was the need for measurements of higher precision than that foreseen by the cartographer. Frisi, on the other hand, was enthusiastic about the project and thought that a map did not require such a high precision. If we take into account the subsequent development of cartography in Italy, as mentioned in the next Sections, which required indeed an even higher precision, the best approach was that of the astronomers. Long and tiring negotiations, however, did not settle the question, owing also to an increasing disagreement between Frisi and the astronomers[11]. In 1778 he published a *eulogium* of Cavalieri (Frisi, 1778*)* containing a harsh criticism of Jesuits and their schools. On their side, in the works published in the *Ephemerides* of the Observatory*,* the astronomers (notably Oriani) cited several authors, but they tended to ignore Frisi, who then criticized the accuracy of their astronomical observations. Sadly, the subsequent documents of accusation and defence and the public expressions, aside from provoking a certain scandal in Milan, exacerbated the situation.

---

[9] See e.g. McConnell (2007; p. 138) and Mandrino et al. (1994). There had been several problems as regards the precision of measurements with the previous large quadrant of Canivet (of 1768).

[10] For the details of the cartography in Brera, see Schiaparelli (1880), Mori (1903), Monti & Mussio (1980), Paolucci et al. (1988).

[11] In the meanwhile Rizzi-Zannoni was working on the map of the territory of Padua. Then from 1781 he was in Naples, where he operated fruitfully.

Only after the death of Frisi, in 1784, the useful project was resumed. In 1786 the astronomers were invited by the authorities to submit a plan for the map, and then to realize it. After the acquisition of the necessary equipment, the operations began in 1788 with the very accurate measurement of the geodetic baseline of Somma (Somma Lombardo), and were continued until 1789 with the triangulation of the lower part of the Duchy, while the mountain region was covered in 1790-1791. In 1792 the material was assembled for the drawings and the preparation of the accurate engraving of copper plates, with the scale 1:86400. The last plate of the *Carta topografica del Milanese e Mantovano* was almost completed in 1796 when the war forced to stop the work. The drawings and the plates were carried away by the Austrian Government; they will be returned to the Observatory by 1804.

## 4. General Bonaparte and Barnaba Oriani in 1796

While the economic, financial and social crisis of the French Revolution of 1789 persisted, the French Republic pursued a military politics in Europe, by disseminating the revolutionary libertarian principles in other countries, and especially by gaining a huge amount of money (war indemnity and revenues) from the States that were 'freed' from the 'oppression' of the Kings. In the spring of 1796, Napoleon Bonaparte (1769-1821), a general not yet 27 years old and very little-known, arrived in Italy in the lead of an ill-equipped army. For a better comprehension, we quote the proclamation of Bonaparte to his army at the beginning of the campaign of Italy (Nice, 27 March 1796): "Soldiers, you are naked, unfed; the government owes you much but has given you nothing. The patience and courage that you have shown among these rocks are admirable; but it brings you no glory—not a glimmer falls upon you. I will lead you into the most fertile plains in the world. Rich provinces, great cities will be in your power; there you will find honour, glory and riches. Soldiers of Italy, do you lack for courage or endurance?"[12] (tr. W. Hanley, 2008; Corresp. 91). It was not by chance that, during the campaign, Bonaparte had therefore to intervene with other proclamations against the pillages[13]; moreover, it is obvious

---

[12] *"Soldats, vous êtes nus, mal nourris; le Gouvernement vous doit beaucoup, il ne peut rien vous donner. Votre patience, le courage que vous montrez au milieu de ces rochers, sont admirables; mais ils ne vous procurent aucune gloire; aucun éclat ne rejaillit sur vous. Je veux vous conduire dans les plus fertiles plaines du monde. De riches provinces, de grandes villes seront en votre pouvoir; vous y trouverez honneur, gloire et richesses. Soldats d'Italie, manqueriez-vous de courage ou de constance?"* As discussed by Hanley (2008), however, this proclamation to the army is questionable at best (and fictitious at worst), and other scholars argued that, while this particular proclamation was never issued to the entire *Armée d'Italie*, something similar might well have been said to the various regiments and demi-brigades individually addressed by the new commanding general.

[13] See, for instance, Hanley (2008; Corresp. 615): *"Ordre du jour. Quartier général, Milan, 23 prairial an IV (11 juin 1796). Le général en chef est informé que, malgré ses ordres réitérés, le pillage continue dans l'armée, et que les maisons des habitants des campagnes sont partout*

that part of the finances of the Duchy of Milan and of other Italian States would have been used for the soldiers' pay and for the benefit of France. In the period between April and May 1796, with rapid manoeuvres, Bonaparte forced the Savoy House (King of Sardinia) to a costly peace in Piedmont, and then he expelled the Austrians from the Duchy[14]. It was this campaign of Italy that revealed suddenly to Europe the great commander. The hectic young general tended to manage his victories in a rather personal way, and that had some implications also for the culture (and astronomy) in Italy.

After the Battle of Lodi[15], on 15 May Bonaparte entered in Milan, and during his short stay, apart from confirming the hard requisitions and taxes (including the first works of fine art to be delivered in France) and setting up a new Government, he met Oriani, most likely following the recommendation of the Directory; the astronomer was in fact considered as the local representative of the *république des lettres*, i.e. the world of science, art and literature[16]. We think that the published historical reconstructions of these facts do not allow a clear understanding of the events concerning Bonaparte and Oriani, and therefore we have attempted to clarify them by taking into account the available documentation.

The French Directory sent a letter to Bonaparte on 16 May (*27 floréal an 4*)[17], where Carnot, president of Directory, recommended to receive and to visit the famous scientists and artists, and, when the General would have taken possession of Milan, to honour and protect particularly the astronomer Oriani, who was very known for his continuing services to the sciences[18]. According to a letter of 22 May (*3 prarial*) addressed to the Directory, Bonaparte had just received the courier "*qui est parti le 26 de Paris*", with the good news of the signed peace treaty with the King of Sardinia (Court of Turin)[19]; that news was contained in a letter dated apparently *27*

---

*dépouillées et dévastées. Cette conduite infâme de la part de quelques individus, qui aspirent au déshonneur et à la perte de l'armée, ne permet plus au général de différer l'emploi des moyens de rigueur qu' il doit déployer pour la conservation de l'ordre public"*.

[14] The governor of the Duchy of Milan was Ferdinand, son of Maria Theresa. After Joseph II, the emperor was his brother Leopold II (1790), and then in 1792 the son of Leopold, Francis II (or I).

[15] 10 May 1796.

[16] In that 'republic', being of noble birth was not important, since the 'citizens' were distinguished only according to the respective merit and achievements.

[17] The letter was published in 1819 in the *Correspondance inédite* of Napoleon (Bonaparte, 1819a, pp. 172-175). A copy was extracted from that publication and preserved in the Archive of the Brera Observatory (AOB); from a comparison with other manuscripts, we can say that the copy was written by Oriani himself (AOB, Fondo Oriani, 214/009).

[18] "*…Le Directoire vous recommande d'accueillir et de visiter les savans et les artistes fameux des pays où vous êtes, et, lorsque vous vous serez emparé de Milan, d'honorer et de protéger particulièrement l'astronome Oriani, si connu pour les services qu'il ne cesse de rendre aux sciences*" (Bonaparte, 1819a, p. 175).

[19] Treaty of Paris of 15 May 1796.

*floréal* (Bonaparte 1819a, p. 176, p. 184)[20]. We think that the courier brought also the other letter that mentioned Oriani, since it had the same date, and after having read it, Bonaparte should have decided to meet the astronomer on 22 May. The respect gained in France by Oriani is remarkable, and it would be interesting to better understand the motivation behind the identification of the astronomer as the representative of the whole Milanese culture. Certainly Lalande was taking care about the Italian astronomers facing the war dangers, as indicated by his short accounts in the *Bibliographie astronomique*[21], and in the next Section we will discuss further this issue. Joseph J.L. de Lalande (1732-1807) had been director of the Paris Observatory (including the Bureau of longitudes) from 1795 to 1800, and member of the Academy of Sciences and of the *Institut National des Sciences et des Arts*. He participated also to the preparation of the *calendrier républicain* and to the standardization of the weights and measures. Presumably, he had therefore to interact with the politicians (Convention, Directory).

After the meeting with Oriani, Bonaparte wrote a famous letter addressed to him, that has been often recalled by historians and scholars who have dealt with the politics and culture of those years. The letter has been published several times and for different reasons, and along with Oriani's reply it has been (mis)interpreted according to the political interests of the time. We think it should be considered just in the context of the events of 1796, when there was the *citoyen* Bonaparte, a young general 'liberator' (or 'oppressor', depending on the point of view), and not yet the great Napoleon. According to a biography of Oriani (Rovani, 1857) the date of the letter was *Milan, 3 prairial* (22 may), while in the *Oeuvres* (Bonaparte, 1822, p. 42) and *Correspondance* (Bonaparte, 1858, p. 322) it was *Milan, 5 prairial* (24 may); note that the general left Milan on 23 May, and went back to the town in the night of 24 May (Bonaparte, 2010). In the letter, Bonaparte proclaimed:

> "The sciences, which ennoble human intelligence, and the arts, which embellish life and transmit great events to posterity, ought to be honored by free governments. All men of genius, all those who have attained a distinguished place in the Republic of Letters, are Frenchmen, whatever may have been their country of birth. In Milan, the learned have not enjoyed the consideration that they should have. Withdrawn in their laboratories, they count themselves lucky if kings and priests do not treat them ill. That is not so today; thought has become free in Italy. No longer is there an inquisition, or intolerance, or despots. Invite the men of learning to meet and to tell me their needs and their views on what should be done to give new life and a new existence to the sciences and fine arts.

---

[20] That is, the courier may have taken about 6 days from Paris to Milan. There could be some misprints or typos as regards the dates: in the letter of *23 prairial* (11 June) of the Directory, the previous one mentioning Oriani is quoted as for *26 floréal*.

[21] Lalande (1803, p. 773) wrote that in 1796 he asked the Directory to recommend to General Bonaparte the astronomers Oriani, De Cesaris and Reggio, and that the General met them, assuring his protection.

All who wish to go to France will be welcomed with honors by the Government. The French attach a greater value to a knowledgeable mathematician, a reputable painter, or a man of distinction of whatever profession, than to the richest and most populous city. Will you then, Citizen, carry these sentiments to the distinguished men of learning in Milan?" [22] (tr. W. Hanley, 2008, Corresp. 491).

This letter was addressed in a certain sense to all the artists and savants of Lombardy, but the audience of the letter extended far beyond the territory of northern Italy; it was reprinted in July 1796 in various Parisian newspapers[23], displaying to the French Bonaparte's sympathies with the arts and sciences. Bonaparte wrote another letter with the same date 24 May to the Municipalities of Milan and Pavia, expressing similar sentiments and encouraging the "celebrated University of Pavia" to resume its normal classes, and also this letter was reprinted in Parisian newspapers (Hanley 2008, Corresp. 492). The young General had a very high consideration of sciences and he considered himself a member of the republic of the letters as a mathematician, however the historians recognize in particular his great skill in propaganda.

Oriani did not receive the letter. He read it one month later, when it was published in a newspaper. We suspect that, in spite of the date, actually it had been written some weeks after the meeting; however, since there is no supporting documentation, ours is just a conjecture. The two letters to Oriani and to the Municipalities of Milan and Pavia probably were *open letters,* and they had been written (or completed) just before 20 June. Our reconstruction is the following.

Bonaparte left from Milan and went to Lodi on 23 May, on the way to Borghetto sul Mincio (south of Garda Lake, in the territory of Venetian Republic, a no belligerent State) where he intended to oppose the Austrian army. Some riots occurred, however, in Milan, and they were subdued by the General Despinoy, who was operating as governor of the town. Revolts occurred also in Pavia, and in the large village of Binasco (midway between Milan and Pavia), so that Bonaparte was forced to come back to Milan[24]. The *Correspondance* (Bonaparte 1819a; 1858) indicates that, after the massacre of Binasco and the repression of Pavia on 26 May, the headquarters were located in the territories of the Venetian Republic, such as Brescia (27-29 May), Valeggio (30 May), Peschiera (31 May - 2 June), Verona and Roverbella (3-5 June), then again Milan (7-11 June), Pavia (12 June), Tortona (13-17 June), Modena (19 June) and finally Bologna (20-26 June) in the Papal States.

---

[22] The following is the version published by Rovani (1857), based on letters preserved after Oriani's death by prof. Lotteri of Pavia: *"Au quartier general. Milan, le 3 prairial. Bonaparte général en chef de l'armée d'Italie au c. Oriani, astronome. Les sciences qui honorent l'esprit humain, les arts qui embellissent la vie, et transmettent les grandes actions à la postérité, doivent être spécialement honorés dans les gouvernemens libres. Tous les hommes de génie, tous ceux qui ont obtenu un rang distingué dans la république des lettres sont Français, quelque soit le pays qui les ait vus naître. Les savans dans Milan n'y jouissaient pas de la considération qu'ils devaient avoir: retirés dans le fond de leur laboratoire ils s'estimaient heureux que les rois et les prêtres voulussent bien ne pas leur faire aucun mal. Il n'en est pas ainsi aujourd'hui; la pensée est devenue libre dans l'Italie; il n'y a plus ni inquisition, ni intolérance, ni despotes. J'invite les savans à se réunir et à me proposer leurs vues sur les moyens qu'il y aurait à prendre, ou les besoins qu'ils auraient pour donner aux sciences, aux arts une nouvelle vie et une nouvelle existence. Tous ceux qui désirent aller en France seront accueillis avec distinction par le gouvernement. Le peuple français ajoute plus de prix à l'acquisition d'un savant mathématicien, d'un peintre de réputation, d'un homme distingué, quel que soit l'art qu'il professe, que de la ville la plus riche et la plus abondante. Soyez donc, Citoyen, l'organe de ces sentiments auprès des savans et artistes distingués qui se trouvent à Milan. Bonaparte"*.

[23] For example in *L'Historien* (Paris), *17 messidor an 4* (7 July 1796) (Hanley 2008).

In the meanwhile, Carnot had written a short letter to Bonaparte dated 11 June (*23 prarial*), asking information about Oriani[25]: "[…] The Directory will know with satisfaction that you fulfilled their indications regarding this distinguished artist [!], and invite you to give an account of what has been done to attest to the citizen Oriani the interest and the estimation that French have had always for him, and to prove that they are able to bring together the love of glory and freedom and that of arts and talents". Assuming that the courier from Paris to the headquarters required about 5 or 6 days, Bonaparte could have received the letter on 16 or 17 June. We think that it was this remind by Carnot that inspired to the General the proclaim to the savants of Milan, represented by Oriani, a proclaim that had to ensue in some way the visit of the astronomer that occurred the previous month, and hence the date had to correspond to the day(s) after that meeting. The open letter to Oriani undoubtedly appears in full agreement with what Carnot had pointed out; Bonaparte sent a copy to the Directory as a demonstration that he had satisfied their requirement (and the letter was then published on Parisian newspapers). This reconstruction makes a sense if we consider that he was a consummate master of public relations.

On 20 June (*2 Messidor*) the Bonaparte's letter appeared on the *Corriere Milanese*, and then it was also affixed at all public places in Milan. Oriani, probably surprised, clearly felt compelled to respond, and that occurred three days later, 23 June (*5 Messidor*) (Gabba, 1929). Meanwhile, on 21 June Bonaparte wrote from

---

[24] The insurrection of Binasco was terribly repressed. There were about two hundred persons killed, the soldiers acted cruelly on the civil population, and, after the pillage, a fire lasting three days (from 24 to 26 May) destroyed half the village. Bonaparte, who was back to Milan in the night of 24 May, was informed about the massacre. In the morning of 26 May he left for Pavia and passed near Binasco. He will recall that impressive sight several times as a terrible example; after the disaster, the village never fully recovered.

[25] AOB, Fondo Oriani 214/009 (see also Bonaparte 1819a, pp. 226-227). *"Le Directoire exécutif, en vous recommandant, par sa lettre du 26 floréal, d'accueillir et de visiter les savans et les artistes fameux des pays dans lesquels vous vous trouvez, vous a désigné particulièrement le célèbre astronome Oriani, de Milan, comme devant être protégé et honoré par les troupes républicaines. Le Directoire apprendra avec satisfaction que vous avez rempli ses intentions à l'égard de cet artiste* [sic] *distingué, et il vous invite en conséquence de lui rendre compte de ce que vous avez fait pour donner au citoyen Oriani des témoignages de l'interêt et de l'estime qu'ont toujours eu pour lui les Français, et lui prouver qu'ils savent allier à l'amour de la gloire et de la liberté celui des arts et des talens."*

Bologna to the Directory one of his reports on the progresses of the campaign of Italy, and he included a brief description of his previous meeting with the astronomer. Oriani did not frequent social gatherings and he was a person having few wants. According to the General (Bonaparte, 1822, pp. 64-65), the astronomer was dazzled by the magnificence of the place, almost certainly the new Serbelloni Palace (restructured and completed in 1793), where Bonaparte was hosted with his general staff, and only after some time he was able to reply. Oriani "finally recovered from his surprise: «Excuse me, he said, but I enter these splendid rooms for the first time; my eyes are not accustomed…» He did not realize that with these few words he criticized bitterly the government of the archduke"[26]. The bitter criticism is just a loose interpretation of the General, as clearly demonstrated by the reply written then by Oriani. The sentence in Bonaparte's letter to the astronomer: "In Milan, the learned have not enjoyed the consideration that they should have. Withdrawn in their laboratories, they count themselves lucky if kings and priests do not treat them ill" is just propaganda, but it could be justified also by the first impression about Oriani, since he appeared *"interdit"* (speechless, dumbfounded), and *"ému, troublé"* (disconcerted), as affirmed by Napoleon few months before his death in St. Helena (see Section 7).

On 23 June Oriani submitted his reply, but the General Despinoy returned it right away asking for some changes, because he deemed it too strong (Rovani, 1857; Gabba, 1929). In this first version he had written that, though he had not received personally the letter published in the *Corriere Milanese* of *2 Messidor* (and that was then affixed at almost all the public places of the town), he thought it was his obligation to reply[27]. Upon the request of Despinoy, he then softened the text and submitted a second version[28], that was sent to Bonaparte on 24 June.

---

[26] *"…J'ai vu, à Milan, le célèbre Oriani: la première fois qu'il vint me voir, il se trouva interdit, et ne pouvait pas répondre aux questions que je lui faisais. Il revint enfin de son étonnement: «Pardonnez, me dit-il, mais c'est la première fois que j'entre dans ce superbes appartemens; mes yeux ne sont pas accoutumés…» Il ne se doutait qu'il faisait par ce peu de paroles, une critique bien amère du gouvernement de l'archiduc. Je me suis empressé de lui faire payer ses appointements et de lui donner les encouragements nécessaires. Vous trouverez ci-joint copie des lettres que je lui ai écrites, dès l'instant que j'ai reçu la recommandation que vous m'avez envoyée pour lui."* (Bonaparte, 1822, pp. 64-65).

[27] *"Quoique je n'aie pas reçu la lettre publiée dans le Courier Milanais le 2 Messidor et qu'on a ensuite affichée à presque tous le endroits publiques de la Ville, je crois qu'il est mon devoir d'y faire deux mots de réponse"*. The manuscript of the two versions of the letter has been reproduced in fac-simile by Gabba (1929); there are, however, some errors of date conversion (or possible misprints) in the comments.

[28] Text from Gabba (1929): *"[Note. La lettre précédent ayant été jugée trop forte par le Général de Division Despinoy, j'en ai changé quelques articles. Après l'avoir corrigée, il m'a promis qu'il l'aurait envoyée au Général en Chef Bonaparte tout de suit le 6 Messidor]. La lettre que vous m'avez fait l'honneur, mon Général, de m'écrire le 3 plairial a été imprimée hier dans le papiers publiques de Milan, je crois donc qu'il est de mon devoir de vous faire une réponse. Si ma façon d'écrire est un peu roide et ne convient pas à la distance qu'il y a entre votre rang de Général en*

Oriani, though recognizing that he was perhaps a bit *"roide"* [stiff], wrote: "Previously, the learned men of Milan were not so despised, nor ignored by the government". Indeed, he added, the scientists had the salary paid by the Austrian Government during the war, but since they arrived, the French had ceased any payment, creating deep consternation in their families. To Bonaparte's exceedingly famous sentence: "All men of genius, all those who have attained a distinguished place in the Republic of Letters, are Frenchmen, whatever may have been their country of birth", Oriani intended to reply: "My General, you do credit to the distinguished savants and artists of all countries by calling them Frenchmen. Actually they belong to the Republic of letters and hence to all Europe, or, better, to the entire world".[29] This phrase, however, was taken off in the second version, along with another 'too strong' sentence: "Since I am convinced that you do not want to earn the men of letters by starving them into submission, I think that the best way to make them love the French Republic is to nourish them"[30]. We could remark, anyway, that Oriani actually represented his colleagues and expressed their needs, but probably not exactly in the way expected by Bonaparte. On the other hand, as regards Oriani himself, he wrote that he had very few needs; moreover, it was just up to him to accept a honourable appointment in one of the most famous European universities with substantial salaries.

Promptly, on 29 June Bonaparte replied from Livorno (Tuscany) that he had 'already' ordered to pay the wages, and thanked Oriani for his care[31].

---

*chef d'une Armée victorieuse et mon état de simple particulier, vous n'avez qu'à déchirer la lettre, je repose entièrement sur les intentions favorables que vous m'avez montrées lorsque j'eu l'honneur de vous parler il y a un mois. Les gens de lettres de Milan n'étaient pas ci devant méprisés ni négligés par le Gouvernement, au contraire ils jouissaient chacun dans sa profession d'une honnête pension et d'une considération proportionnée à leur mérite. Dans la guerre actuelle, quoique très dispendieuse, tous les appointemens ont été payés régulièrement chaque mois, et ce n'est que depuis quelques semaines que tout payement a cessé, et qu'on ignore même quand il recommencera. Dans plusieurs familles des gens de lettres il y a une vraie consternation sur le manque absolu de subsistance pour le présent et pour l'avenir. Il me semble que l'unique moyen de faire cesser leurs calamités et de leur inspirer de l'affection pour la République Française soit de les nourrir, en donnant ordre au Caissier de l'Instruction Publique de leur payer tout de suite les appointemens du mois de Mai passé, et ceux de juin qui va finir. J'espère que le Général en chef voudra bien attribuer ces sentimens à l'amour que j'ai pour la vérité et la justice. Car en mon particulier, ayant très peu de besoins, je saurai vivre en quelque pays que ce soit, et d'ailleurs dans ce moment même il ne dépend que de moi d'accepter une charge honorable dans une des plus célèbres Universités de l'Europe avec des appointemens considérables. De l'Observatoire de Brera, le 5 messidor de l'an 4,$^e$ de la République Française une et indivisible. Oriani astronome de Milan"*.

[29] *"Vous faites, mon Général, l'honneur aux savans et aux artistes distingués de tous les pays de les appeler Français. Réellement ils appartiennent à la République littéraire et par conséquent à toute l'Europe ou pour mieux dire à toute la terre"*.

[30] *"Comme je suis persuadé que vous ne voulez pas gagner les gens de lettres par la famine, je crois que le meilleur moyen de leur faire aimer la République Française soit de les nourrir ..."*

## 5. General Bonaparte and the astronomers

After the Battle of Arcole (15-17 November 1796), nearby Verona, against the Austrian Army, Bonaparte was back to Milan at the beginning of December 1796. On 5 December, he wrote a short (and a bit flattering) letter to Lalande as a reply, informing him that he had quickly forwarded the letter (of Lalande) addressed to the astronomer of Milan (maybe Oriani). He added some enthusiastic opinions about science, and in particular astronomy: "among all the sciences, astronomy has been the most useful to reason and to commerce". We think that this statement is remarkable, since it shows that at the end of the Age of Enlightenment people were aware of the fundamental importance of astronomy for the cultural and economic progress. Moreover, he wrote that it is especially astronomy that "needs far communications", i.e. communications at the international level (we could use the expression: astronomical globalization), and "the existence of the republic of letters". The young General then expressed an unusual opinion about the work of an astronomer: "To share the night between a woman and a beautiful sky, and spend the day putting together observations and calculations, it seems to me the happiness on Earth" (Bonaparte, 1859a, p. 138)[32].

Lalande had occasion to intervene on Italian astronomers' behalf in April 1797, after the bombardment of Verona, in the territory of the Republic of Venice. The Republic was a no belligerent country, but Verona was an important place of transit for Austrian and French armies. In Verona there was a private observatory built by Antonio Cagnoli (1743-1816), with several instruments. Since Lalande was taking care of this astronomer as well as those in Milan, he certainly sent letters concerning him. In fact, in August 1796 Bonaparte had written to General Augereau recommending to respect the person and property of Cagnoli[33]. Unfortunately, during the bombardment of Verona ("*Pasque Veronesi*"), his house was hit and the observatory was damaged. Bonaparte was informed by Lalande[34], and in July 1797 he ordered to compensate the astronomer for this damage, and moreover to give funds for the Italian Society of Sciences, whose President was Cagnoli. This Society had been founded specifically as *Italian* even though a State called *Italia* did not exist yet[35], and its members were forty learned men of the States and Statelets in which the country was then divided. After the treaty of Campo Formio of October 1797, the Republic of Venice ceased to exist, and part of its territories were ceded to Austria in exchange for the recognition of the Cisalpine Republic, proclaimed in July 1797. Verona was ceded later, in January 1798, but Bonaparte, in any event, took a very quick decision: on 7 November 1797 he decreed the transfer of the Italian Society of Sciences from Verona to Milan, capital of the Cisalpine Republic, and, consequently, the transfer of Cagnoli, that was appointed astronomer of Brera with the same salary of Oriani. Later he ordered to purchase the astronomical instruments of Cagnoli, part for Brera and part for Bologna Observatory, and granted annual funds for the Italian Society[36]. Cagnoli was appointed also member of the "*Corpo Legislativo*" (i.e. the Parliament) of the Cisalpine Republic, but he felt very uncomfortable with this appointment, and in April 1798 he accepted the teaching of mathematics at the new Military Academy in Modena, although that implied a lower salary and leaving Brera[37]. He will stay in Modena until 1807, and this town then became the effective site of the Society, that enjoyed the granted annual funds for several years.

---

[31] *"Au Quartier Général. Livorno, le 11 messidor, A. 4ᵉ de la République une et indivisible. Au citoyen Oriani astronome. J'avais déjà donné les ordres pour que les savans, qui étaient pensionnés, continuassent à toucher leurs appointemens. Je réitère les mèmes ordre aussi aux agens militaires de Milan que je vous prie de voir, afin que cela ne souffre aucun retard dan l'exécution. Je vous remercie de la peine que vous vous êtes donnés de me prévenir des besoins qu'avaient vos collègues. Je n'oublierai rien pour les faire cesser. Je suis avec estime et considération. Bonaparte"* (Rovani, 1857; Gabba, 1929).

[32] *"J'ai reçu, Citoyen, votre lettre du 28 octobre. Je me suis empressé de faire passer celle qui était incluse pour l'astronome de Milan. Toutes les fois que je puis être utile aux sciences et aux hommes qui les cultivent avec autant de succès, je suivis mon inclination, et je sens que je m'honore. De toutes les sciences, l'astronomie est celle qui a été la plus utile à la raison et au commerce; c'est surtout celle qui a le plus besoin de communications lointaines et de l'existence de la république des lettres; heureuse république, où les hommes sont souvent, comme dans les autres, en proie aux passions et à l'envie, mais où la gloire est accordée au mérite et au génie, qui l'obtiennent sans partage! Partager une nuit entre une jolie femme et un beau ciel, et le jour, le passer à rapprocher ses observations et ses calculs, me paraît être le bonheur sur terre".*

An account written by Cagnoli in 1799 and addressed to the Austrian chancellor Thugut adds further information to what reported in the available official documents[38]. The astronomer wrote (in French) that Bonaparte had become fixated ("*s'entêta*") on the transfer of the Italian Society. Cagnoli had made a list of very reasonable and practical motivations in order to convince Bonaparte to keep the Society in Verona, but "*la volonté du plus fort prononcée si fortement ni admettait point de choix*", the will of the mighty expressed so mightily did not allow alternatives. Probably, the General was strongly interested in its national, Italian character, given his attempt to build a sort of Italian State. In fact, on 9 November 1797, just two days after the decree concerning the Society, Bonaparte decided the establishment of the National Institute in Bologna[39] (*Istituto Nazionale*; see for example http://www.istitutolombardo.it/storia.html). The Institute, committed to art and science, had been included in the Constitution of the Cisalpine Republic of July 1797 (the article is very similar to that of the Constitution of 1795 of the French Republic regarding its *Institute National*), however we think it was not by chance that the decrees concerning the two scientific entities, the Society and the Institute, were promulgated very near in time.

---

[33] AOB, Fondo Cagnoli, 201/024. Antonio Cagnoli had begun to study astronomy in Paris with Lalande at age 37.

[34] Lalande (1803, p. 792).

[35] The Society was founded in Verona by A. Lorgna in 1782. Presently its name is *Accademia Nazionale delle Scienze detta dei XL*, and its site is Rome. For an account about the Society, in particular during the Napoleonic epoch, see Cagnoli (1802).

[36] Letters and documents in AOB, Fondo Cagnoli, 201/027 /031 /032 e 042. The decrees have been published in Bonaparte (1859b, p. 406, p. 437).

[37] According to Cagnoli, leaving from the Observatory was a greater sorrow than the decreased salary. However, it may be possible that the arrival of a new astronomer in Brera, with his own ideas, had been considered a difficulty by Oriani, Reggio and DeCesaris (e.g. Dal Prete 1998).

[38] Austrians were back to Italy on the summer 1799, and the Cisalpine Republic had ceased to exist.

In April 1798, while he was preparing the campaign of Egypt, Bonaparte wrote from Paris to General Brune recommending him to protect the Observatory of Milan and in particular Oriani, who had complained about the bad behaviour against him[40]. During this period, in fact, the astronomers in Milan had to face difficulties with the local Government of the Cisalpine Republic. The astronomers and the other employees had to swear eternal hate for the Government of the Kings. On 23 April Oriani had written to the Cisalpine Directory that he refused to swear: "Barnaba Oriani respects all the well organized governments, but he does not understand why, in order to observe planets and stars, does one need to swear eternal hate for one government or the other"[41]. He added that he had been hired at age 23 at the Specola of Brera by a monarchic government, and that he got some reputation in his profession thanks to the resources assigned by the same government during 20 years; therefore he would have been the most ungrateful man if he swore hate for who did just good for him. Then he declared to submit to the law and to leave the Observatory. The Cisalpine Directory tried privately to convince the astronomers Reggio, Cesaris and Oriani to change their decision, but without success; eventually, a compromise was reached and the astronomers swore simply obedience to the laws (Rovani, 1857; Gabba, 1929). In spite of the interventions of Bonaparte, the astronomers had again to face hassle, when some local anticlericalists tried to demonstrate their own 'valour' with the three priest astronomers[42].

---

[39] As we will see in the next Section, the Institute became really operative only after 1804 (with the Kingdom of Italy); in 1810 the then Royal Institute was moved to Brera. The Constitution of Cispadane Republic (that consisted of some territories of the region Emilia-Romagna: Bologna, Ferrara, Reggio Emilia, Modena) of March 1797 already contained an article about it, mentioning also its importance for the improvement of agriculture.

[40] "*Je vous recommande de protéger l'observatoire de Milan, et, entre autres, Oriani, qui se plaint de la conduite que l'on tient à son égard: c'est le meilleur géomètre qu'il y ait eu*" (Bonaparte, 1819b, p. 56). Probably the General had been informed by Lalande.

[41] "*Barnaba Oriani, astronomo della specola di Brera, stima e rispetta tutti i governi bene ordinati, né sa comprendere come per osservare le stelle ed i pianeti sia necessario giurare odio eterno a questo od a quel governo. Egli è stato in età di 23 anni impiegato nella Specola di Brera da un governo monarchico, e si acquistò qualche nome in questa professione coi mezzi che gli vennero dal medesimo governo accordati per 20 anni continui. Egli sarebbe dunque il più ingrato degli uomini se ora giurasse odio a chi non gli ha fatto che del bene*" (Rovani, 1857). We note that the local government was probably not 'well organized'.

## 6. From the Republic to the Kingdom

In spring-summer 1799, while Bonaparte was in Egypt, the Austrian-Russian army defeated the French army in Northern Italy and the Cisalpine Republic was dissolved. Bonaparte restored it in the summer of 1800 (Battle of Marengo), and after the Treaty of Lunéville (9 February 1801) the territory of the Republic was extended on western and eastern sides. The Brera Observatory did not suffer of such overturning, given probably the renowned firm position of Oriani with respect to politics and governments.

During the second period of the Republic, Oriani was appointed Count and Senator, and in 1802 he participated at the Council of Lyon, where the Italian Republic replaced the Cisalpine one, and Bonaparte was 'freely' elected President. In the same year he gave to Oriani a new astronomical clock (Arnold, an English instrument); according to Lalande (1803, p. 793), it was more sophisticated than the existing instruments, with pivots which rotated on rubies, anchor escapements encrusted with diamonds and a compensator made of iron and zinc: a true masterpiece.

F. Melzi d'Eril, Vice-President of the Italian Republic, decreed in 1802 that the topographic maps of the Milanese territory had to be redrawn, given that the engravings were lacking, and in any case the available material had to be corrected and extended to the other provinces not included in the ancient Duchy of Milan. New instruments were bought for Brera, and, starting again from the geodetic baseline of Somma, triangulations were performed with higher precision than in 1788. Schiaparelli (1880) wrote that the work was not as successful as the previous one, and the campaign of 1804 was especially fatal, when almost all operators fell sick, one person drowned in a river, and Reggio died due to excessive labour. The works proceeded too slowly. In 1801 a Deposit of War had been instituted in Milan, in which the materials for the topographical maps were collected and preserved; later, it was changed by the Austrians into the Military Geographic Institute (*Istituto Geografico Militare*). Engineers of the Deposit were already working on a map of Italy as a continuation of that of France (Cassini), and they collected all the astronomical-geodetic surveys carried out until that time. In 1805 they requested to Brera the measurements already made by the astronomers, in order to extend the triangulations. Since it did not make much sense to redo the same work, and given the evident difficulties, in 1808 it was decided to stop the activity of Brera on the topography. Mori (1903) quoted a worthy document written by Cesaris in that occasion, where the merits of the astronomers of the Observatory had been recalled, particularly the education of many engineers, that had become engineers of the Deposit. Mori then remarked the importance of the Brera school of geodetic measurements and land surveying, since it would have had a large and beneficial influence on the topographic activity performed in the first half of the century in many Italian regions.

---

[42] On February 1799, Oriani wrote to the Ministry of Police of the Republic complaining "*le continue vessazioni*", the continuous vexations suffered by him and the other employees of Brera. According to a "military" proclaim, not only the astronomers had to continue paying a tax even though there was an exemption according to the law, nay, "*i così detti preti devono pagare il doppio per essere celibi*", the so-called "priests" had to pay twice because they were unmarried (AOB, Fondo Oriani, 214/011).

During the 19$^{th}$ century the astronomers made other specific geodetic works[43]. For instance, at the request of Laplace, Oriani had been in charge of a new measurement of the arc of meridian between Rimini and Rome, more than fifty years after the one performed by Boscovich and Maire (Tagliaferri and Tucci, 1993); however, the previous results were essentially confirmed. The reason for the Laplace's request was related to the problem of the figure of the Earth, since the assumed ellipsoid was only partly in agreement with observations. For example, both Boscovich and G.B. Beccaria had been among the first scientists that remarked the gravitational effect of the mountains on the geodetic measurements.

After his proclamation as Emperor of the French, Napoleon changed the Italian Republic into Kingdom. On 17 March 1805 he was proclaimed King of Italy, and in June Eugène de Beauharnais (child by first marriage of Josephine, then wife of Napoleon) was named Viceroy. The *Istituto Nazionale* had been established officially in 1802 in Bologna, and it had been divided into three sections: physical and mathematical sciences, political and moral sciences, literature and arts. It was considered the main body of the culture in the State, and it was an important reference for all that concerned the public education and the university. It became operative few years later. In February 1808 the Brera Palace was declared Palace of the Sciences and the Arts, and at the end of 1810 it became the new site of the *Istituto*, that was named *Istituto Reale di Scienze, Lettere ed Arti*, with sections also in Venice, Bologna, Padua and Verona. One of the results of the suppression of many churches and convents (not only in Milan) during the period of the Republic and the Kingdom was the destination of many paintings to Brera, increasing considerably in this way the patrimony of the *Pinacoteca*. The new museum opened officially in 1809[44].

During this period, Oriani refused the appointment as Ministry of Public Education, but he was in charge of the reorganization of the Universities of Pavia and Bologna, and gave contributions to the activity of the *Istituto* (Bianchi, 1933). From 1802 (to 1859) the Brera Observatory and its school of astronomy were joined to the University of Pavia. Moreover, Oriani had been appointed to the chair of the

---

[43] Also the geodesy school of astronomers will continue. In the second half of 19$^{th}$ century G.V. Schiaparelli and G. Celoria will teach geodesy in the newly founded Polytechnic of Milan.
[44] By the way, it may be worth recalling that also the music conservatory of Milan had been established by Napoleon in 1807.

Commission for the introduction of the decimal metric system. Napoleon offered him the bishopric of Vigevano, but he refused; however, he accepted an annual pension related to that title, in order to continue more easily his profession as an astronomer[45].

In 1801, after the discovery of Ceres at the Observatory of Palermo by Giuseppe Piazzi (1746-1826), the asteroids and their orbits were studied also at Brera Observatory, with Francesco Carlini (1783-1862) as new astronomer. In 1813 he began to work with Giovanni Plana (1781-1864) of Turin on a theory of the Moon, a work that got a prize by the French Academy of Sciences in 1820. Given the favourable cultural environment of the Kingdom, an interesting activity in optics took also place, which unfortunately could not be continued after the Restoration. In 1811 Giovan Battista Amici (1786-1863)[46], a young teacher from Modena, submitted to the Government two metal mirrors and a microscope to be examined by experts. The astronomers tested the mirrors and concluded that the quality and the performances were similar to those of their Herschel telescope, and proposed to the Government to order a larger mirror to Amici, so that they would have been no more compelled to buy perfect instruments abroad. Then a sort of competition began: Giovanni G. Gualtieri (? -1852) from Modena, former collaborator of Amici, sent to the Government a large mirror (25 cm) to be examined. The astronomers gave their approval and the mirror was purchased by the Government. In November 1811 Amici provided his new 30 cm mirror, which was evaluated positively by the astronomers. The reflectors had very long focal lengths, and that was a problem since there was not a large enough place where to install them. In 1812 Gualtieri offered a first positive solution with a reduced focal length. In the subsequent years the astronomers tried to get funds to realize a complete telescope and to install it in a new site, but it was a too expensive project[47]. Then in 1814 the Austrians were back to Italy; unfortunately, they had a very different attitude towards culture than twenty years earlier.

## 7. The Restoration

In 1814-1815, after the defeat of Napoleon, the European Restoration fought against the ideas of the revolution spread by the French armies, and the absolutist kings were re-established wherever possible. In 1814 the Kingdom of Italy ceased to exist, and with it also the hopes for independence[48]. The new *Regno Lombardo-Veneto*

---

[45] Given Oriani's modest lifestyle, as discussed in the next Section, this and other money will be very important for the Milanese culture after his death.

[46] Amici will be appointed director of Arcetri Observatory in Florence (founded in 1807) in 1832.

[47] For further details about the archive documents on the Amici telescope, see Broglia and Antonello (2005).

[48] A sufficiently shared idea of independence was actually lacking in Milan. The *austriacanti* (pro-Austrians) wanted just the previous regime; the French party was related to Melzi d'Eril and Eugène of Beauharnais; the *italici puri* (pure Italians) were against France, and later they deceived themselves trying to get the independence for free by diplomacy; the *murattiani,* the followers of

(Kingdom of Lombardy-Venetia) was under the Austrian rule. When the Emperor of Austria Francis I visited Milan at the beginning of 1816 he did not seem to be interested in the progresses of science and even in meeting Oriani[49]. In 1817, at age 65, the astronomer officially left his position, but he continued to live for some years in the Specola and do research. He died in 1832. In the Anniversary Address to the Royal Society, the Duke of Sussex expressed the following opinion about Oriani: "Upon the whole, if the union of practical with theoretical science be considered, we shall be justified in pronouncing him to have been, after Bessel, the most accomplished astronomer of the present age" (obituary reported by Urban, 1833).

Oriani bequeathed a huge amount of money to people and institutions; for example, to Plana for his final work on the theory of the Moon, and to the Ambrosian Library, where he got an help when he was a young student[50]. With the money for the Observatory it was possible to fill the positions of "second astronomer" and of "third student" for many years until the Italian unification (1861). According to Schiaparelli (1880), we have reason to believe, that, without that will of Oriani, the Observatory would not exist anymore or it would have been abandoned as it had been the case of several others in Italy. Oriani handed down also an important legacy. His professional experience, along with that of Reggio and Cesaris, was passed on to students and young researchers that had the opportunity to be educated or to specialize in Brera: they were most part of the leading Italian astronomers of the first decades of 19[th] century.

Before concluding, let me go back for a while to Napoleon. The former-Emperor of the French was in exile in St. Helena, and yet there he had the occasion to recall Oriani and his first meeting with him. His impression was still vivid in 1820, when he was seriously ill, just few months before his death. It is worth quoting the physician Antommarchi (1826, English edition, pp. 365-368).

> "14th [November 1820] Napoleon again took up the subject of Italy, and spoke at length of Oriani. «He is the greatest geometrician that ever existed!» Napoleon had treated him with great distinction, protected him, and

---

Murat, King of Naples, who betrayed Napoleon and was trying to reach an agreement with Austrians, supported him in the place of Beauharnais.

[49] According to some patriots (e.g. Misley, 1832), during the visit to the Institute (now named *Imperial Regio Istituto del Lombardo-Veneto*), the Austrian emperor would have ignored ostensibly Oriani and declared: *"Signori, non domando loro scienza; non domando che religione e moralità"*, "Sirs, I do not ask you science, but just religion and morality". The emperor visited again Brera in 1825. That time he promised to take care of the astronomers' needs, and indeed a new meridian circle was provided. It was realized by Starke at the Imperial-Royal Polytechnic Institute of Vienna (founded in 1815), but for various reasons it was not the instrument desired by the astronomers (see Schiaparelli 1880; Antonello 2010).

[50] A bust of Oriani is displayed proudly among the benefactors of the Ambrosian Library (*Biblioteca Ambrosiana*). The Library was founded in the 16th century. It preserves old (from 2nd century AD on) and rare manuscripts and codices (such as the *Codex Atlanticus* of Leonardo da Vinci). There is also an art gallery (*Pinacoteca*).

recommended him to Brune when he went out on his expedition to Egypt. He had taken pleasure in testifying publicly the respect he entertained for his learning, by writing to him immediately after he had entered Milan; thus honouring in his person all those who cultivate sciences in Italy:     [the text of the letter follows].

Napoleon had preserved a most particular recollection of this celebrated man; he often spoke of him, and took pleasure in relating the details of the first audience he had given him. He described Oriani's timidity and embarrassment at the sight of the stately retinue of the staff, which quite dazzled him, and the trouble he had to restore to him confidence and composure.  «You are here with your friends; we honour learning, and only wish to show the respect we entertain for it – Ah! General, excuse me, but this pomp and splendour quite overpower me; I am not accustomed to witness them». He, however, recovered his self-possession, and held with Napoleon a long conversation, which produced in his mind a feeling of surprise which he could not for a long time overcome. He was unable to conceive how it was possible to have acquired, at the age of twenty-six, so much glory and science: the General was to him an inexplicable phenomenon".[51]

Napoleon's ambition to be a scientist is remarkable; we do not know Oriani's opinion about this account of Bonaparte, which was published in 1825. Anyway, from the Antommarchi's diary one gets the impression that the remind of the astronomer was a bit of relief for the sufferings of the 'last days'.

**Conclusion**

In 18th century the Brera Observatory contributed actively not only to the astronomical research but also to the development of the Duchy of Milan, and its importance reached a peak during the Napoleonic epoch. There was a certain cultural liveliness in the years after 1760, that became 'effervescence' by the end of the century, which was a much troubled period, but that showed the interesting potentialities of an Italian State, even though it was limited to Northern Italy and under the French control. The capital of this State was Milan, and Napoleon probably intended to strengthen the multicultural character of Brera Palace, the main cultural centre of the town, where both the sciences and the arts were present. The astronomer Barnaba Oriani gave significant contributions to the realization of that idea. He was a highly respected personality in Italy and Europe, and his firm position certainly played some role in those troubled days; the available documents suggest moreover a particular respect by Napoleon, in spite of the unusual beginnings of their acquaintance.

---

[51] Antommarchi was the physician (actually he was an anatomist) of Napoleon from September 1819 until his death. During the exile, Napoleon received books and newspapers from European friends and from his family, that were important for writing his *Mémoires* (with the main help of Bertrand, Montholon, Gourgaud and Las Cases; see e.g. Bonaparte, 2010). Among the last books there were the volumes of his *Correspondance inédite*, (e.g. Bonaparte, 1819a, b); as we noted in Sect. 4, they were read also by Oriani in Italy. Napoleon relived the events mentioned in the letters. *"Napoléon avait conservé un souvenir tout particulier de ce savant célèbre [Oriani], et se plaisait à revenir sur les détails de la première audience qu'il avait donnée. Il le voyait encore ému, troublé, ébloui par l'appareil de l'état majeur. Il avait eu beaucoup de peine à le calmer. «Vous êtes au milieu de vos amis; nous honorons le savoir» … Il se remit cependant, et eut avec Napoléon une longue conversation qui le jeta dans un étonnement dont il fut bien plus long-temps à revenir. Il ne concevait pas comment à vingt-six ans on pouvait avoir acquis tant de gloire et de science. Le général était pour lui un phénomène inexplicable"*.